\documentclass[a4paper,11pt]{article}
\usepackage{styleBuding} 
\usepackage{multirow}
\usepackage{lineno}
\usepackage{booktabs}
\usepackage{siunitx}
\usepackage[table]{xcolor}
\usepackage{xcolor}

\newcommand{\cts}[1]{\textcolor{violet}{[Citation needed]}}

\title{\boldmath Exploring vector-like $B$-quark pair production at CLIC in fully hadronic final states}

\author[a,b]{Baoxia Wang}
\author[a,b]{Shuo Yang,}
\author[c]{and Pengxuan Zhu\footnote{Corresponding author}}
\affiliation[a]{School of Physics and Electronic Technology, Liaoning Normal University, Dalian, 116029}
\affiliation[b]{Center for Theoretical and Experimental High Energy Physics, Liaoning Normal University, Dalian, 116029}
\affiliation[c]{ARC Centre of Excellence for Dark Matter Particle Physics \& CSSM, Department of Physics, Adelaide University, Adelaide, SA 5005}

\emailAdd{shuoyang@lnnu.edu.cn}
\emailAdd{baoxiawang1022@163.com}
\emailAdd{pengxuan.zhu@adelaide.edu.au}

\abstract{
We investigate the discovery potential of the $3~{\rm TeV}$ Compact Linear Collider (CLIC) for a singlet vector-like bottom partner $B$ decaying via $B \to tW$. Focusing on the fully hadronic final state $B\bar{B} \to tW\,tW$, we reconstruct boosted top and $W$ candidates using large-$R$ Valencia jets, supplemented by a merging strategy for partially resolved decays. A systematic scan of the jet-radius parameter identifies $R=0.8$ as the optimal choice, balancing boosted-jet containment with jet multiplicity. Using a cut-based analysis optimized for the $(2t+2W)$ topology and an integrated luminosity of $5~{\rm ab}^{-1}$, CLIC can achieve sensitivity to $m_B \lesssim 1.5~{\rm TeV}$. These results highlight CLIC’s excellent capability to probe heavy vector-like quarks in high jet multiplicity environments, extending well beyond the reach of current hadron collider searches.
}

\begin{document}
\maketitle
\flushbottom

\section{Introduction}
\label{sec:intro}
After the discovery of the $125~{\rm GeV}$ Standard Model (SM) Higgs boson at the Large Hadron Collider (LHC)~\cite{ATLAS:2012yve, CMS:2012qbp}, addressing the electroweak hierarchy (naturalness) problem remains a central goal of particle physics. A wide range of theories, including supersymmetry~\cite{Dimopoulos:1981au,Dimopoulos:1981zb,Witten:1981nf,Haber:2017aci, Martin:1997ns}, composite Higgs models~\cite{Cacciapaglia:2021uqh, Kaplan:1983fs, Agashe:2004rs} and little Higgs constructions~\cite{Arkani-Hamed:2001nha, Arkani-Hamed:2002ikv, Schmaltz:2005ky}, stabilize the Higgs mass and often predict new states at the TeV scale.

\par Vector-like quarks (VLQs) arise naturally in many of these frameworks. Because their left- and right-handed components transform identically under the SM gauge group, a gauge-invariant mass term is allowed, independent of electroweak symmetry breaking. Such states play key roles in naturalness, collider phenomenology and electroweak precision tests~\cite{Aguilar-Saavedra:2013qpa, Branco:2022gja, Banerjee:2022izw}. They also appear in scenarios of spontaneous CP violation and non-axionic solutions to the strong CP problem~\cite{tHooft:1976rip, tHooft:1976snw, Nelson:1983zb, Branco:1983tn, Nelson:1984hg, Barr:1984qx, Bento:1991ez, Branco:2003rt}, can assist gauge-coupling unification~\cite{Gursey:1975ki, Choudhury:2001hs, Mohapatra:2006gs, Dermisek:2012ke}, and have been discussed in relation to the first-row CKM unitarity tension via mixing with an up-type VLQ~\cite{Seng:2018yzq, Belfatto:2019swo, Belfatto:2021jhf, Botella:2021uxz, Branco:2021vhs, Alves:2023ufm, Crivellin:2022rhw}.

\par In a model-independent language, third-generation VLQs can appear as ${\rm SU}(2)_L$ singlets ($T$ or $B$), doublets [$(X,T)$, $(T,B)$, $(B,Y)$], or triplets [$(X,T,B)$, $(T,B,Y)$], with mixing to SM quarks via Yukawa interactions, leading to diverse collider signatures. In particular, the vector-like bottom partner $B$ and the top partner $T$ have been extensively studied at hadron colliders~\cite{Yue:2005kv, Gong:2019zws, Li:2019tgd, Gong:2020ouh, Benbrik:2024fku, Banerjee:2024zvg, Aguilar-Saavedra:2017giu, Das:2018gcr, Benbrik:2019zdp, Cacciapaglia:2019zmj, Aguilar-Saavedra:2019ghg, Zhou:2020byj, Wang:2020ips, Corcella:2021mdl, Cui:2022hjg, Banerjee:2022xmu, Bhardwaj:2022wfz, Bhardwaj:2022nko, Bardhan:2022sif}. The ATLAS and CMS collaborations have reported numerous model-independent searches for single and pair production of VLQs~\cite{ATLAS:2024fdw, ATLAS:2022hnn, CMS:2024bni, ATLAS:2022tla, ATLAS:2023ixh, CMS:2018kcw, CMS:2019eqb, CMS:2020ttz, CMS:2022fck, CMS:2024xbc}. For the vector-like $B$ decaying into SM modes $B\to tW$, $bZ$ and $bH$, current limits typically reach the $\mathcal{O}(1.2$--$1.3)\,{\rm TeV}$ scale for singlet- and doublet-like scenarios~\cite{ATLAS:2025tdo, CMS:2024bni, Benbrik:2024fku}. These bounds, however, rely on the assumption that only these three channels are present with fixed branching fractions; in non-minimal settings where $B$ can decay into beyond-the-SM states, constraints can be substantially relaxed~\cite{Banerjee:2023upj, ATLAS:2024itc}.

\par The fully hadronic $B\bar B\to tW\,tW$ channel is of particular interest. At the LHC it suffers from overwhelming QCD multijet backgrounds, large combinatorics in high jet multiplicity, and pile-up, all of which degrade jet-substructure observables and complicate data-driven background estimation~\cite{Yang:2014usa, Choudhury:2021nib}. In contrast, the \emph{Compact Linear Collider} (CLIC) offers a clean $e^+e^-$ environment with well-defined initial states and substantially reduced QCD activity, which is advantageous for boosted-object reconstruction and multi-jet final states~\cite{Buchkremer:2013bha, Franceschini:2019zsg, Qin:2021cxl, Han:2021kcr, Han:2021lpg, Qin:2022mru, Han:2022exz, Han:2022zgw, Han:2022rxy, Yang:2023wnv, Qin:2023zoi, Han:2024xyv, Yang:2025ktj, Marzani:2019hun, Bardhan:2025pmn}.

In this work we investigate a singlet vector-like bottom partner $B$ at the $3~{\rm TeV}$ stage of CLIC, focusing on the fully hadronic $B\bar B\to tW\,tW$ final state. We use large-$R$ Valencia jets to reconstruct boosted $t$ and $W$ objects and a simple merging strategy to recover partially resolved decays. The simplified model setup is presented in Sec.~\ref{sec:model}. The boosted-jet reconstruction and tagging definitions are detailed in Sec.~\ref{sec:jet}. The search strategy, background composition and sensitivity estimates are given in Sec.~\ref{sec:collider}. We conclude in Sec.~\ref{sec:sum}.

\section{\label{sec:model}Vector-like $B$ model preliminaries}
\begin{figure}[th]
\centering
\includegraphics[width=0.5\linewidth]{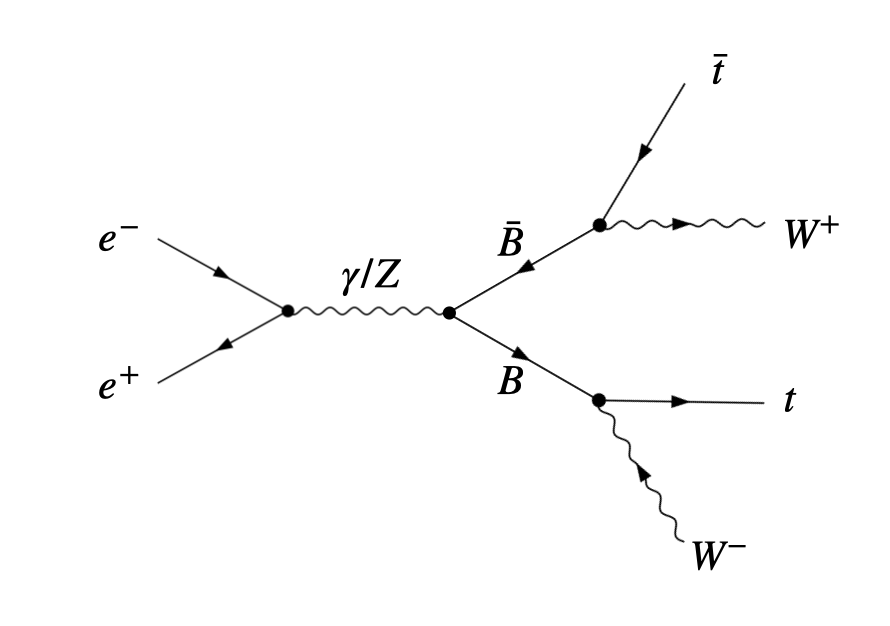}
\caption{\label{fig:diam} Representative diagram for $e^+e^-\to B\bar B$ with subsequent decays $B\to tW$, $bZ$, and $bH$.}
\end{figure}
Vector-like quarks (VLQs) are fermions whose left- and right-handed components transform identically under the SM gauge group, so their mass term is gauge invariant and independent of electroweak symmetry breaking. In this work we consider a bottom-like partner $B$ with electric charge $-1/3$ that is an ${\rm SU}(2)_L$ singlet and mixes predominantly with the third generation. This choice is both theoretically well-motivated and phenomenologically minimal for collider studies.

At energies relevant for CLIC, the interactions of a singlet $B$ with third-generation quarks and electroweak bosons can be captured by the following effective Lagrangian
\begin{equation}\begin{split}
	\mathcal{L} &= \bar{B} (i D_\mu \gamma^{\mu} - m_B) B  \\ 
	&+ \kappa \left( \frac{g}{\sqrt{2}} \bar{B}_L W_\mu^- \gamma^\mu t_L 
	+ \frac{g}{2 \cos{\theta_W}} \bar{B}_L Z_\mu \gamma^\mu b_L 
	- \frac{m_B}{v} \bar{B}_R H b_L 
	\right) + L \leftrightarrow R \\
	&+ {\rm h.c.}
\end{split}\end{equation}
Here $P_{L,R}=(1\mp\gamma_5)/2$, $c_W \equiv \cos\theta_W$, $v\simeq246~{\rm GeV}$, and $\kappa$ is an overall coupling normalisation that scales the partial widths. We neglect couplings to light generations. Results presented below depend primarily on branching fractions rather than the absolute value of $\kappa$, because $B\bar B$ pair production at an $e^+e^-$ collider proceeds via $s$-channel $\gamma/Z$ exchange and is fixed by the electroweak charges of $B$.

For a pure singlet, the partial widths satisfy the well-known relation
$\Gamma_{tW}:\Gamma_{bZ}:\Gamma_{bH}\approx 2:1:1$, leading to the approximate branching ratios
\begin{equation}
{\rm BR}\bigl(B\to tW\bigr)\simeq 50\%,\qquad 
{\rm BR}\bigl(B\to bZ\bigr)\simeq 25\%,\qquad 
{\rm BR}\bigl(B\to bH\bigr)\simeq 25\%.
\end{equation}
If additional beyond-the-SM channels $B\to X$ are open, then the SM modes then obey
\begin{equation}\label{eq:brs2}
\begin{split}
{\rm BR}(B \to tW) &= \tfrac{1}{2}\bigl(1 - {\rm BR}(B \to X)\bigr), \\
{\rm BR}(B \to bZ) &= {\rm BR}(B \to bH) = \tfrac{1}{4}\bigl(1 - {\rm BR}(B \to X)\bigr).
\end{split}\end{equation}
Unless otherwise stated, our baseline results assume the singlet pattern with ${\rm BR}(B \to X)=0$.

At CLIC, $B\bar B$ pairs are produced through $e^+e^-\!\to\!\gamma/Z\!\to\!B\bar B$, as illustrated in Fig.~\ref{fig:diam}. The corresponding cross section is determined by $m_B$, the centre-of-mass energy, and the electroweak couplings of $B$; it is insensitive to the decay parameters $\kappa$ apart from their impact on the final-state composition. In the following we take $(m_B,\, {\rm BR}(B\to tW))$ as the two parameters that span the simplified-model interpretation.

\section{\label{sec:jet}Fat jets at CLIC}
\begin{table}[t]
\centering
\resizebox{0.98\linewidth}{!}{
\begin{tabular}{
    r c@{\hspace{10mm}}  S[table-format=2.1] S[table-format=2.1] S[table-format=2.1] S[table-format=2.1] c@{\hspace{8mm}} 
    S[table-format=2.1] S[table-format=2.1] S[table-format=2.1] S[table-format=2.1]
}
\toprule
\multirow{2}{*}{$N_{W\text{-jets}}$} &&
\multicolumn{4}{c}{ \sf BP1: $m_{B} = 1.2~{\rm TeV}$} && \multicolumn{4}{c}{ \sf BP2: $m_{B} = 1.45~{\rm TeV}$} \\
\cmidrule(lr){3-6} \cmidrule(lr){8-11} 
&&  {$R=0.8$} & {$R=1.0$} & {$R=1.2$} & {$R=1.5$} &&
    {$R=0.8$} & {$R=1.0$} & {$R=1.2$} & {$R=1.5$} \\
\midrule
\multicolumn{2}{c}{Total $\geq 4$ jets}
& 79.97 & 42.35 & 30.44 &  16.07 
&& 76.75 & 35.57 &  24.70 &  12.29 \\
[1ex]
\multicolumn{11}{c}{\bf Subgroup: $N_{t\text{-jets}} = 2$} \\ 
$\geq4$ && 0.03 & 0.00 & 0.00 & 0.00 && 0.02 & 0.00 & 0.00 & 0.00 \\
3 &&  0.31 & 0.01 & 0.00 & 0.00 && 0.28 & 0.01 & 0.00 & 0.00 \\
2 && 12.19 & 2.54 & 1.32 & 0.38 && 10.10 & 1.86 & 0.91 & 0.23 \\
1 && 2.63 & 1.82 & 1.24 & 0.51 && 2.56 & 1.35 & 0.81 & 0.28 \\
0 && 6.96 & 2.48 & 1.68 & 0.80 && 5.58 & 1.59 & 1.02 & 0.42 \\[0.5ex]
Subtotal && 22.12 & 6.85 & 4.24 & 1.69 && 18.54 & 4.81 & 2.74 & 0.93 \\
[1ex]
\multicolumn{11}{c}{\bf Subgroup: $N_{t\text{-jets}} = 1$} \\
$\geq4$ && 0.06 & 0.01 & 0.00 & 0.00 && 0.06 & 0.01 & 1.02 & 0.00 \\
3 && 1.20 & 0.53 & 0.27 & 0.08 && 1.48 & 0.46 & 0.23 & 0.06 \\
2 && 25.25 & 11.75 & 7.42 & 2.99 && 24.47 & 8.77 & 5.05 & 1.69 \\
1 && 1.58 & 1.10 & 0.74 & 0.34 && 1.60 & 0.84 & 0.52 & 0.21 \\
0 && 7.60 & 5.43 & 4.40 & 2.61 && 7.52 & 4.62 & 3.42 & 1.82 \\[0.5ex]
Subtotal&& 35.69 & 18.82 & 12.83 & 6.02 && 35.13 & 14.70 & 10.24 & 3.78 \rm\\
[1ex]
\multicolumn{11}{c}{\bf Subgroup: $N_{t\text{-jets}} = 0$} \\
$\geq4$ && 0.16 & 0.07 & 0.04 & 0.01 && 0.21 & 0.06 & 0.03 & 0.01 \\
3 && 1.77 & 0.88 & 0.47 & 0.14 && 1.94 & 0.69 & 0.35 & 0.09 \\
2 && 16.93 & 11.43 & 8.60 & 4.54 && 17.07 & 9.43 & 6.54 & 3.05 \\
1 && 0.20 & 0.18 & 0.15 & 0.10 && 0.22 & 0.21 & 0.16 & 0.09 \\
0 && 3.09 & 4.11 & 4.09 & 3.55 && 3.65 & 5.65 & 5.65 & 4.36 \\[0.5ex]
Subtotal && 22.15 & 16.67 & 13.35 & 8.34 && 23.09 & 15.99 & 12.73 & 7.51\\
\bottomrule
\end{tabular}
}
\caption{\label{tab:vjamb} Reconstruction performance of the Valencia jet algorithm for $B\bar B$ at two benchmarks, $m_B=1.2~{\rm TeV}$ (BP1) and $m_B=1.45~{\rm TeV}$ (BP2), as a function of jet radius $R$. All entries are percentages of the total generated signal. The first row gives the preselection efficiency for requiring $N_{{\rm jets}}\ge4$. The remaining rows split the sample by the number of top- and $W$-tagged jets; the three `Subtotal' rows sum to the first row (up to rounding) for each benchmark and $R$}
\end{table}

In this section, we discuss the reconstruction of $W$-jets and top-jets, and clarify the effect of jet radius $R$ on jet reconstruction efficiency. As illustrated in Fig.~\ref{fig:diam}, the full decay chain
$B\to t(\to bW)\,W \to b\,j\,j\,j\,j$ produces up to five jets per $B$ quark. For TeV-scale $B$ quarks at CLIC, the large $m_B$ typically yields boosted top and $W$ objects, whose hadronic decay products can merge into a single ``fat'' jet.

We employ the Valencia sequential-recombination algorithm~\cite{Boronat:2014hva, Boronat:2016tgd} to cluster large-$R$ jets. The algorithm defines the pairwise and beam distances
\begin{equation}
\begin{split}
 d_{ij} &= \min \big(E_i^{2\beta},E_j^{2\beta}\big)\,\frac{1-\cos\theta_{ij}}{R^{2}},\\
 d_{iB} &= E_i^{\,2\beta}\,\sin^{2\beta}\theta_i\, ,
\end{split}
\end{equation}
with tunable energy-weighting parameter $\beta$ and effective cone size $R$. We work with $\beta=1$ and scan $R\in[0.8,1.5]$, which interpolates between better containment of boosted objects (larger $R$) and better resolution of nearby jets (smaller $R$).

The Valencia algorithm is designed for future CLIC simulation studies. By varying $R$ within the range $[0.8, 1.5]$, it can effectively capture both narrow light-quark jets and highly boosted "fat" jets resulting from heavy-resonance decays.
In this work, a top-tagged ($W$-tagged) jet is defined as a Valencia jet with a reconstructed mass within $\pm 30~{\rm GeV}$ ($\pm 20~{\rm GeV}$) of the top-quark ($W$-boson) mass. We first require at least four Valencia jets; events that do not meet this preselection are vetoed. For events with fewer than two top jets, a recursive process will attempt to identify additional top candidates. Firstly, the all jet candidates are sorted by their masses. Next, the jet pair with the largest mass is combined with the remaining light jets, and the combination whose invariant mass is closest to the top quark mass is selected as the top candidate. Jets selected by the top candidate are removed from the jet list, and this process continues until all possible top candidates have been chosen. If the combined number of final top candidates and top jets exceeds two, only the top candidate whose mass is closest to the top quark mass is retained. The other top candidates are no longer kept, and the paired jets are returned to the jet list. Similarly, a procedure will be applied to jets with mass less than the $W$ mass to construct additional $W$ candidates. 
Finally, we count the number of top-tagged jets, $N_{t\text{-jets}}$, and $W$-tagged jets, $N_{W\text{-jets}}$, and summarized in Table~\ref{tab:vjamb}. 
\par In Table~\ref{tab:vjamb}, all entries are expressed as percentages of the total generated signal sample for each benchmark and jet radius. The first row ("Total $\geq4$ jets") represents the preselection efficiency for requiring at least four Valencia jets. The subsequent sections further categorize the selected events by $(N_{t\text{-jets}}, N_{W\text{-jets}})$. Within each section, the ``Subtotal'' indicates the fraction of events with a fixed $N_{t\text{-jets}}$ after preselection. For any given combination of benchmark and jet radius $R$, the three subtotals (corresponding to $N_{t\text{-jets}} =2$, $=1$, and $=0$) sum to the value in the first row, up to rounding.

From Table~\ref{tab:vjamb}, two notable features can be observed:
\begin{itemize}
  \item The behaviour of the preselection efficiency with varying jet radius $R$.
  \par Requiring $N_\text{jets} \ge 4$ naturally penalises large radii, because nearby partons are more often clustered into a single fat jet. For BP1 ($m_B=1.2~{\rm TeV}$) the preselection efficiency falls from $79.97\%$ at $R=0.8$ to $ 16.07\%$ at $R=1.5$, and for BP2 ($m_B=1.45~{\rm TeV}$) it drops from ${76.75\%}$ to ${12.29\%}$. This steady decline argues for smaller $R$ values when maintaining jet multiplicity is essential.

  \item The efficiency of top and $W$ reconstruction. 
  \par Fully hadronic top jets are more difficult to tag than $W$ jets because their three-prong structure must include both a $b$ sub-jet and a $W$ sub-jet. At $R=0.8$, the $N_{t\text{-jets}}=1$ category is the most populated after preselection (BP1: ${35.69\%}$; BP2: ${35.13\%}$), while $N_{t\text{-jets}}=2$ is less common (BP1: ${22.12\%}$; BP2: ${18.54\%}$), and $N_{t\text{-jets}}=0$ is comparable (BP1: ${22.15\%}$; BP2: ${23.09\%}$). Increasing $R$ further suppresses the $N_{t\text{-jets}}=2$ category, as distinct sub-jets are more likely to be merged.
  \end{itemize}
  
Balancing boosted-object containment with the need to maintain four resolved jets, we adopt $R=0.8$ as our default. This choice maximises the preselection efficiency and retains sizeable rates in the $N_{t\text{-jets}} = 2$ and $N_{W\text{-jets}} = 2$ categories used by the cut-based analysis in next section.

\section{\label{sec:collider}Detection potential for vector-like $B$ at CLIC}
We now turn to the expected sensitivity of the $3~{\rm TeV}$ CLIC programme to the fully hadronic signal $B\bar B\to tW\,tW$. After the boosted-object reconstruction described in Sec.~\ref{sec:jet}, only SM processes that can mimic four energetic fat jets with two masses near $m_t$ and two near $m_W$ remain relevant.
\begin{figure}[th]
\begin{center}
\includegraphics [width=0.49\linewidth] {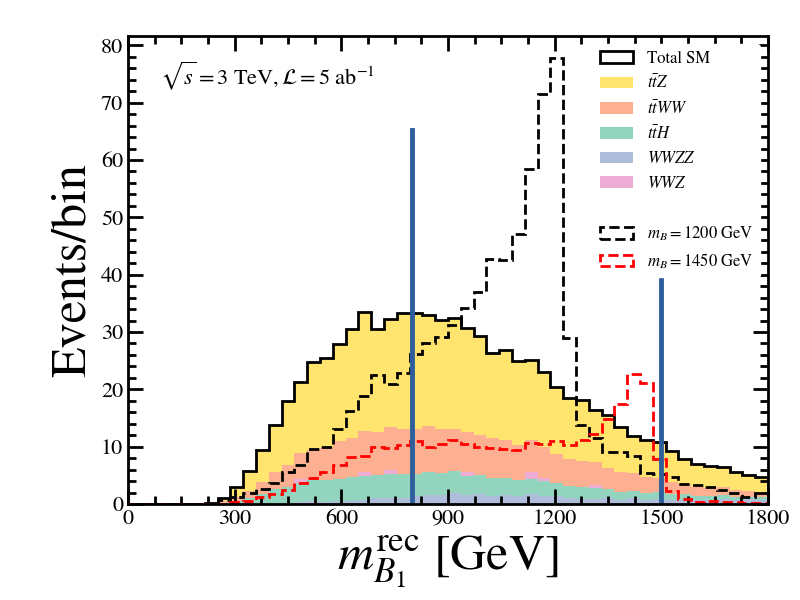}
\includegraphics [width=0.49\linewidth] {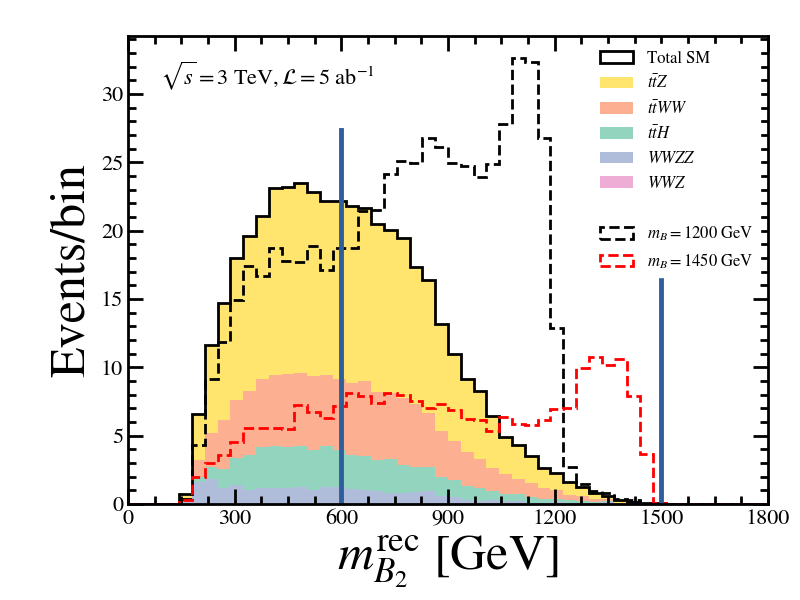}
\caption{The reconstructed invariant mass distributions for the two $B$ candidates after applying Cut-5 and Cut-6. The left panel shows the heavier candidate $m_{B_1}^{{\rm rec}}$ and the right panel the lighter $m_{B_2}^{{\rm rec}}$. Stacked histograms represent the total SM background (solid fill) and its main components, while dashed curves correspond to the benchmark signals with $m_B = 1.2~{\rm TeV}$ (black) and $m_B = 1.45~{\rm TeV}$ (red). Vertical lines indicate the chosen SR mass windows. Both distributions are normalised to $5~{\rm ab}^{-1}$ at $\sqrt{s}=3~{\rm TeV}$.}
\label{fig:jetmassbh}
\end{center}
\end{figure}

\begin{table}[th]
\centering
\resizebox{\linewidth}{!}{
\begin{tabular}{
l c@{\hspace{5mm}} 
S[table-format=1.6] 
S[table-format=1.6] 
S[table-format=1.6]
S[table-format=1.6]
S[table-format=1.6]
S[table-format=1.6]
S[table-format=2.2] c@{\hspace{5mm}}
S[table-format=1.6]
S[table-format=2.2]
S[table-format=2.1] 
S[table-format=1.6]
S[table-format=2.2]
S[table-format=2.1]
}
\toprule
\multirow{2}{*}{\textbf{Cuts}} 
&& \multicolumn{7}{c}{\textbf{SM backgrounds}} 
&& \multicolumn{3}{c}{\textsf{BP1}: $m_B = 1.2~{\rm TeV}$} & \multicolumn{3}{c}{\textsf{BP2}: $m_B = 1.45~{\rm TeV}$} \\ 
\cmidrule(lr){3-9} \cmidrule(lr){11-13} \cmidrule(lr){14-16}  
&& {$\sigma(t\bar{t}WW)$} & {$\sigma(t\bar{t}H)$} & {$\sigma(t\bar{t}Z)$} & {$\sigma(WWZ)$} & {$\sigma(WWZZ)$} & {$\sigma({\rm SM})$}& {$\varepsilon$ [\%]} 
&& {$\sigma$} & {$\varepsilon$ [\%]} & {$\mathcal{Z}_{A}$} & {$\sigma$} & {$\varepsilon$ [\%]} & {$\mathcal{Z}_{A}$} \\
\midrule
No cuts 
&& 0.3363 & 0.4085 & 1.652  &  32.86   & 1.213 & 36.47    & {--} 
&& 1.074 & {--} & 12.5 & 0.5049 & {--} & 5.9\\
Cut-1
&& 0.2613 & 0.2513 & 0.9816 & 3.394     & 0.4646 & 5.352    & 14.6 
&&  0.8590 & 85.9  & 25.6 & 0.3875 & 76.8  & 11.7\\
Cut-2
&& 0.0256 & 0.2505 & 0.9448 & 3.169     & 0.4412 & 4.831    & 13.3 
&& 0.8587 & 85.9  & 26.9 & 0.3875 & 76.7  & 12.3\\
Cut-3
&& 0.1847 & 0.2199 & 0.7149 & 2.065     & 0.3315 & 3.516    & 9.6 
&& 0.7912 & 79.1  & 28.8 & 0.3545 & 70.2  & 13.2\\
Cut-4
&& 0.0782 & 0.0820 & 0.2910 & 0.0729    & 0.0292 & 0.553    & 1.5 
&& 0.2597 & 26.0  & 23.1 & 0.1047 & 20.7  & 9.7\\
Cut-5
&& 0.0389 & 0.0213 & 0.1005 & 0.0094    & 0.0113 & 0.1816   & 0.50 
&& 0.1600 & 16.0  & 23.6 & 0.0641 & 12.7  & 10.1\\
Cut-6
&& 0.0213 & 0.0102 & 0.0512& 0.0002   & 0.0014 & 0.0843   & 0.23 
&& 0.1219 & 12.2 & 25.0 & 0.0476 & 9.4   & 10.7\\
Cut-7
&& 0.0109 & 0.0005 & 0.0256 & 0.0000   & 0.0006 & 0.0376   & 0.10
&& 0.0864 & 8.6  & 24.8 & 0.0356 & 7.1   & 11.5\\
\bottomrule
\end{tabular}
}
\caption{\label{tab:cutflow}
Cut flow for $B\bar{B}\to tW\,tW$ at $3~{\rm TeV}$ CLIC with ${\rm BR}(B \to tW)\approx 0.5$. All values of $\sigma$ represent cross sections in fb after each selection step. The statistical significance $\mathcal{Z}_A$ is calculated using the Asimov formula for a reference luminosity of $\mathcal{L}=5~{\rm ab}^{-1}$. The relative efficiency $\varepsilon$ is defined with respect to the previous step.}
\end{table}

The irreducible backgrounds arise from multi-boson and top-associated production:
\begin{itemize}
    \item $e^+e^-\to t\bar t\,W^+W^-$, 
    \item $e^+e^-\to t\bar t H$ with hadronic $H$ decays, 
    \item $e^+e^-\to t\bar t Z$ with hadronic $Z$ decays, 
    \item $e^+e^-\to W^+W^-Z$, 
    \item $e^+e^-\to W^+W^-ZZ$.
\end{itemize}
The second category of reducible backgrounds, pure QCD multi-jet events, could potentially contaminate the signal region. However, the four-jet QCD rate at CLIC is already low, and the additional requirement that two jets fall within the top-quark mass window and two within the $W$-boson mass window further suppresses these backgrounds to a negligible level.

In practice, parton-level events are generated using \texttt{MadGraph5\_aMC@NLO 3.3.2}~\cite{Alwall:2014hca} and subsequently showered and hadronized with \texttt{Pythia~8}~\cite{Sjostrand:2014zea}. Detector effects are simulated with the CLIC card in \texttt{Delphes}~\cite{deFavereau:2013fsa, Leogrande:2019qbe}. Jet clustering is performed using \texttt{FastJet}~\cite{Cacciari:2011ma} with the Valencia algorithm~\cite{Boronat:2014hva,Boronat:2016tgd} and a radius parameter of $R=0.8$, unless stated otherwise. Fat jets are required to have $p_T \geq 25~{\rm GeV}$ and $|\eta| < 2.5$. Detailed top- and $W$-tagged jets are defined in Sec.~\ref{sec:jet}.  Finally, a cut-based analysis is performed using \texttt{MadAnalysis-5}~\cite{Conte:2012fm}.

We define a single signal region using the following sequence of selection criteria. Each step is designed to suppress a specific class of background events while preserving the characteristic $2t+2W$ topology. The exclusion significance, $\mathcal{Z}_A$, for a counting experiment with an expected signal $S$ on top of a known background $B$, is used to estimate the expected discovery significance~\cite{Cowan:2010js},
\begin{equation}
	\mathcal{Z}_{A} = \sqrt{2\left( (S+B)\ln\left( 1+\frac{S}{B}\right) - S\right)}. 
\end{equation}
The cut-flow procedure is outlined below and summarized in Table~\ref{tab:cutflow}:
\begin{itemize}
	\item Cut-1: require at least four Valencia jets; veto events with isolated electrons or muons with $p_{\ell} > 10~{\rm GeV}$ is removed.
	\item Cut-2: the signal event is characterized by a large jet multiplicity. Therefore, the transverse momentum of the jet system, defined as
		\begin{equation}
			H_{\rm T} = \sum_{i=1}^{4} p_{\rm T}(j_i),
		\end{equation}
		is required to be greater than $500~{\rm GeV}$ to suppress background events. 
	\item Cut-3: the leading jet $p_{\rm T}$ is required to exceed $400~{\rm GeV}$. As shown in Table~\ref{tab:cutflow}, this criterion significantly suppresses the multi-boson backgrounds $WWZ$ and $WWZZ$.	
	\item Cut-4: exactly two top jets are required, i.e., $N_{\text{t-jet}} = 2$.
	\item Cut-5: one or two $W$-jets are required. For events with $N_{W\text{-jet}} = 1$, an additional jet with a mass close to that of the $W$ boson is considered as the second $W$ candidate.
	\item Cut-6: In this step, we pair the two top-jets with the two $W$-jets to reconstruct two $B$ candidates. Of the two possible pairings, we select the one that minimizes the difference in invariant mass between the candidates. The heavier $B$ candidate is labeled $B_1$, and the lighter one is labeled $B_2$. The distribution of the reconstructed $B_1$ mass, denoted as $m_{B_1}^{\rm rec}$, after Cut-5 is shown for both benchmark scenarios and the background in the left panel of Fig.~\ref{fig:jetmassbh}. Both benchmarks exhibit clear peaks in $m_{B_1}^{\rm rec}$, while the Standard Model background is flatter and extends beyond $\sqrt{s}/2$. In our analysis, we require $800~{\rm GeV} \leq m_{B_1}^{\rm rec} \leq 1500~{\rm GeV}$. 
	\item Cut-7: $600~{\rm GeV} \leq m_{B_2}^{\rm rec} \leq 1500~{\rm GeV}$. As shown in the right panel of Fig.~\ref{fig:jetmassbh}, after applying Cut-6, the $m_{B_2}^{\rm rec}$ distribution for the signal still peaks around $m_B$, while the majority of the SM background is concentrated below $900~{\rm GeV}$. To retain more signal events, we adopt a wider selection range of $[600, 1500]~{\rm GeV}$.
\end{itemize}
From Table~\ref{tab:cutflow}, it can be seen that the SM backgrounds are efficiently suppressed after applying the signal event selection cuts. The statistical significance $\mathcal{Z}_A$ for the signal reaches 24.8 for $m_B = 1.2~{\rm TeV}$ and 11.5 for $m_B = 1.45~{\rm TeV}$, respectively.

\begin{figure}[htb]
\begin{center}
\includegraphics [width=0.49\linewidth] {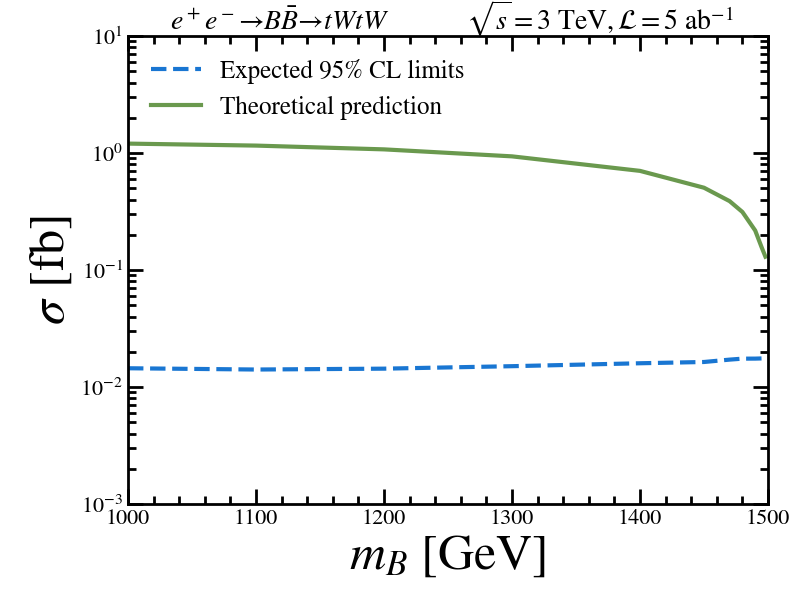}
\includegraphics [width=0.49\linewidth] {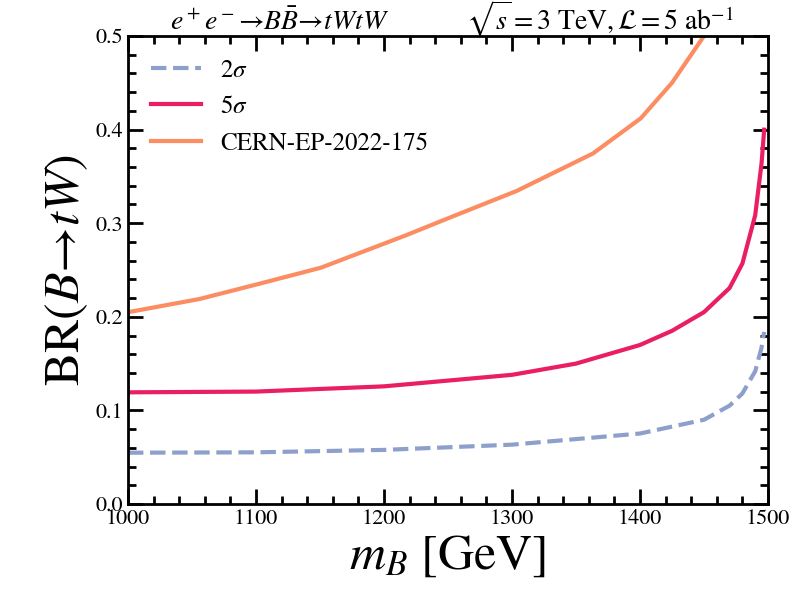}
\caption{
Expected exclusion and discovery reach for the process $e^+e^-\!\to\!B\bar{B}\!\to\!tW\,tW$ at $\sqrt{s}=3~{\rm TeV}$ with an integrated luminosity of $5~{\rm ab}^{-1}$.
The left panel shows the expected $95\%$ confidence level upper limit on the total cross section (blue dashed) compared with the theoretical prediction for a singlet vector-like $B$ quark (green solid).
The right panel displays the corresponding $2\sigma$ exclusion (blue dashed) and $5\sigma$ discovery (red solid) contours in the $(m_B,\,{\rm BR}(B\!\to\!tW))$ plane, together with the current LHC constraint (orange curve, Ref.~\cite{CMS:2022fck}).}
\label{fig:limit}
\end{center}
\end{figure}

Finally, the expected detection limits are presented in Fig.~\ref{fig:limit}. The left panel displays the $95\%$ confidence level (C.L.) upper limits, along with the theoretical prediction for the process $e^+ e^- \to B\bar{B} \to tWtW$. The expected $95\%$ C.L. upper limit on the total production cross section reaches approximately $10^{-2}~{\rm fb}$ across the 1 to 1.5 TeV mass range, which is well below the predicted signal rate for a singlet $B$. As a result, the entire parameter space up to $m_B \lesssim 1.5~{\rm TeV}$ remains within the exclusion reach of CLIC.
The right panel displays the expected $95\%$ C.L. exclusion and $5\sigma$ discovery contours in the $(m_B, {\rm BR}(B\to tW))$ plane. 
For ${\rm BR}(B\to tW)=0.5$, both $5\sigma$ discovery and $2\sigma$ exclusion limits are expected up to about $1.5~{\rm TeV}$.
For scenarios with a large new physics branching ratio, ${\rm BR}(B\to X)$, CLIC achieves a competitive reach compared to the current LHC constraints from Ref.~\cite{CMS:2022fck}, surpassing them by nearly a factor of four in mass coverage.

These results highlight the advantages of the linear collider’s clean environment and high energy, which allow for precise reconstruction of multi-jet final states and significantly expand the search potential for heavy vector-like quarks beyond what is possible at hadron colliders.

\section{\label{sec:sum}Conclusion and outlook}

\par In this study we have evaluated the prospects for discovering pair-produced singlet vector-like $B$ quarks at the $3~{\rm TeV}$ stage of CLIC through the fully hadronic $B\bar{B}\!\to\!tW\,tW$ final state.  
The analysis exploits the Valencia jet algorithm for boosted-object reconstruction and demonstrates that a jet radius of $R=0.8$ provides the best compromise between signal acceptance and object separation in high-multiplicity events.  

\par After a sequence of optimized selection cuts, the Standard-Model backgrounds are reduced to the sub-femtobarn level while preserving signal efficiencies at the $10\%$ level.  
The results indicate that with $5~{\rm ab}^{-1}$ of integrated luminosity, CLIC can achieve both a $5\sigma$ discovery and a $95\%$ C.L. exclusion for $m_B \lesssim 1.5~{\rm TeV}$, significantly extending the current reach of the LHC.

\par These findings highlight the unique potential of future high-energy $e^+e^-$ colliders for probing heavy vector-like quarks in complex multi-jet final states, benefiting from a clean environment and precise detector performance.  
Future work may incorporate systematic uncertainties, jet-substructure taggers, and alternative decay channels ($B \to bZ$, $bH$, or exotic states) to establish a fully model-independent interpretation of vector-like quark searches at CLIC.

\acknowledgments
This work was supported in part by the National Natural Science Foundation of China under Grants No. 12147214 and No. 12575106,  and the Basic Research Project of Liaoning Provincial Department of Education for Universities under Grants  No.~LJKMZ20221431 and Teaching Reform Research Project for graduates of Liaoning Normal University. We would like to acknowledge Xiaoran Zhou for her contributions in the early stages of this work. P. Zhu is supported by the ARC Discovery Project DP220100007.

\bibliography{references}
\bibliographystyle{CitationStyle}

\end{document}